\documentclass[preprintnumbers, prd, onecolumn, showpacs,floatfix,preprintnumbers,
superscriptaddress, nofootinbib]{revtex4}
\usepackage{graphicx}
\usepackage{epsfig}
\usepackage{bm}
\usepackage{amssymb}
\usepackage{float}
\usepackage{amsmath}
\usepackage{subfigure}
\usepackage{dcolumn}
\usepackage{cancel}
\usepackage[colorlinks]{hyperref}
\usepackage[usenames,dvipsnames]{color}
\hypersetup{
     breaklinks=true,
    pdfstartview={FitH},    % fits the width of the page to the window
    colorlinks=true,       % false: boxed links; true: colored links
    linkcolor=blue,          % color of internal links
    citecolor=red,        % color of links to bibliography
    filecolor=magenta,      % color of file links
    urlcolor=blue,           % color of external links
    anchorcolor=green,      % Color for anchor text
    linktocpage=true
}

\def\doi{http://doi.org}

\begin{document}
%%%%%%%%%%%%%%%%%%%%%%%%%%%%%%%%%%%%%%%%%%%%%%%%%%%%%%%%%%%%%%%%%%%%%%%%%%%%%%%%%%%%
\title{A new parametrization of dark energy equation of state leading to double exponential potential}
%%%%%%%%%%%%%%%%%%%%%%%%%%%%%%%%%%%%%%%%%%%%%%%%%%%%%%%%%%%%%%%%%%%%%%%%%%%%%%%%%%%%%%%%%%%%%%%%%%%%%%
%%%%%%%%%%%%%%%%%%%%%%%%%%%%%%%%%%%%%%%%%%%%%%%%%
\author{Sudipta Das}
\email{sudipta.das@visva-bharati.ac.in}
\affiliation{Department of Physics, Visva-Bharati, Santiniketan -731235, India}
%%%%%%%%%%%%%%%%%%%%%%%%%%%%%%%%%%%%%%%%%%%%%%%%%%%%%%%%%%%%%%%%%%%%%%%%%%%%%%%%%%%%%%%%%%%%%%%
\author{Abdulla Al Mamon}
\email{abdulla.physics@gmail.com}
\affiliation{Department of Mathematics, Jadavpur University,Kolkata-700032, West Bengal, India} 
%%%%%%%%%%%%%%%%%%%%%%%%%%%%%%%%%%%%%%%%%%%%%%%%%%%%%%%%%
\author{Manisha Banerjee}
\email{banerjee.manisha717@gmail.com}
\affiliation{Department of Physics, Visva-Bharati, Santiniketan -731235, India} 
%%%%%%%%%%%%%%%%%%%%%%%%%%%%%%%%%%%%%%%%%%%%%%%%%%%%%%%%%%%%%%%%
\pagestyle{myheadings}
\newcommand{\be}{\begin{equation}}
\newcommand{\ee}{\end{equation}}
\newcommand{\bea}{\begin{eqnarray}}
\newcommand{\eea}{\end{eqnarray}}
\newcommand{\bc}{\begin{center}}
\newcommand{\ec}{\end{center}}
%%%%%%%%%%%%%%%%%%%%%%%%%%%%%%%%%%%%%%%%%%%%%%%%%%%%%%%%%%%%%%%%%%%%%%%%%%
\begin{abstract}
We show that a phenomenological form of energy density for the scalar field can provide the required transition from  decelerated ($q>0$) to accelerated expansion ($q<0$) phase of the universe. We have used the latest Type Ia Supernova (SNIa) and Hubble parameter datasets to constrain the model parameters. The best fit values obtained from those datasets are then used to reconstruct $\omega_{\phi}(z)$, the equation of state parameter for the scalar field. The results show that the reconstructed forms of $q(z)$ and $\omega_{\phi}(z)$ do not differ much from the standard $\Lambda$CDM value at the current epoch. Finally, the functional form of the relevant potential $V(\phi)$ is derived by a parametric reconstruction. The corresponding $V(\phi)$ comes out to be a double exponential potential which has a number of cosmological implications. Additionally, we have also studied the effect of this particular scalar field dark energy sector on the evolution of matter over-densities.  
\end{abstract} 
\pacs{98.80.Hw\\
Keywords: Cosmic acceleration, Quintessence field, Dark energy density, Perturbation}
\maketitle
%%%%%%%%%%%%%%%%%%%%%%%%%%%%%%%%%%%%%%%%%%%%%%%%%%%%%%%%%%%%%%%%%%%%%%%%%%%%%%
\section{Introduction}\label{intro}
%%%%%%%%%%%%%%%%%%%%%%%%%%%%%%%%%%%%%%%%%%%%%%%%%%%%%%%%%%
Recent cosmological observations (Riess 1998, 2004; Perlmutter 1999; Eisenstein 2005; Spergel 2007) strongly suggest that the universe is currently going through an accelerated phase of expansion. These observations also suggest that the observed accelerated phase is indeed a very recent phenomenon and the universe must be having a decelerated phase of expansion in the past in order to facilitate the structure formation of the universe. The driving force responsible for generating this observed accelerated expansion is popularly named as dark energy (DE) which has large negative pressure. For review on DE models, one can refer to the relevant review works (Sahni \& Starobinsky 2000; Copeland et al. 2006; Martin 2008). Amongst the most popular DE models, the $\Lambda$CDM model enjoys more worthy attention in the literature, which is found to be in good agreement with the observational data. But, it has two associated theoretical problems, namely, {\it fine tuning problem} and {\it cosmological coincidence problem} (Weinberg 1989; Steinhardt 1999). Alternatively, quintessence models do not suffer from the above mentioned problems due to their dynamical nature and are widely used as candidate for DE. The quintessence (or canonical) field, is capable of driving the acceleration with some suitably chosen potentials, but none of the models have firm theoretical motivation [for a comprehensive review, see (Sahni \& Starobinsky 2000)]. Numerous DE models have been explored and studied over the last two decades in order to explain this observed late time accelerated behaviour of the universe (for details, one can look into (Sahni 2004)]. But, none of these models can be considered as superior to others, so the search is still on for a suitable model for DE consistent with the current observations. Although it is mostly believed that DE components do not cluster, recently studies
are being made to see the effects of perturbations on DE components (Weller \& Lewis 2003; Bartolo 2004; Unnikrishnan 2008a; Unnikrishnan et al. 2008b; Jassal 2010). So this branch of cosmology requires huge attention to probe whether such clustering can provide us with new
information regarding the true nature of the DE component.\\
%%%%%%%%%%%%%%%%%%%%%%%%%%%%%%%%%%%%%%%%%%%%%%%%%%%%%%%%%%%%%%%%%%%%%%%%%%%%%%%%%%
\par Keeping in mind the above facts, we have proposed a simple scalar field model of DE in the framework of a spatially flat ($k=0$) FRW universe, where we have considered a functional dependence for the energy density of the scalar field, $\rho_{\phi} (a)$. The aim of this paper is to investigate the evolution history of the universe in this scenario. With this input, the expressions for the Hubble parameter $H(z)$, the deceleration parameter $q(z)$, the equation of state parameter $\omega_{\phi}(z)$ and the density parameter $\Omega_{\phi}(z)$ are found out. Next, we have obtained the constraints on various parameters of the model using the SNIa, Hubble and joint analysis of SNIa+Hubble datasets. The best-fit values obtained are then used to constrain the evolution behaviors of $q(z)$ and $\omega_{\phi}(z)$. We have found that for this specific ansatz, the deceleration parameter $q$ smoothly transits from the positive to the negative value regime at the recent past (around $z<1$) such that the structure formation can take place unhindered. This results are compatible with those results as expected both theoretically (Padmanabhan 2003; Choudhury 2005) and observationally (Riess 2001, 2004; Turner \& Riess 2002; Chuna 2009; Mamon \& Das 2016). We have also discussed about the future evolution dynamics of the universe. Using the combination of the SNIa and Hubble datasets, we have also tried to obtain the functional dependence of the potential $V(\phi)$ for this model. Finally, we have also looked at the effect of this particular DE sector on the growth of matter perturbations by comparing it with well studied cosmological models such as  $\Lambda$CDM model and CPL model or with a model where there is no DE sector. \\
%%%%%%%%%%%%%%%%%%%%%%%%%%%%%%%%%%%%%%%%%%%%%%%%%%%%%%%%%%%%%%%%%%%%%%%%%%%%%%%%%%%%%
\par The organization of the paper is as follows. In section \ref{sec2}, we have presented the basic equations related to the scalar field dark energy model for a spatially flat FRW model of the universe. We have then obtained analytical solutions for the field equations using a specific choice of $\rho_{\phi}$. In section \ref{data}, we have described the observational datasets and their analysis method used in this paper. We have then obtained the constraints on the various cosmological parameters. The results are presented in section \ref{result}. In section \ref{perturbation}, we have studied the effect of this particular DE sector on the evolution of matter over-densities at perturbative level. Finally, some conclusions are presented in section \ref{conclusion}.
%%%%%%%%%%%%%%%%%%%%%%%%%%%%%%%%%%%%%%%%%%%%%%%%%%%%%%%%%%%%%%%%%%%%
%%%%%%%%%%%%%%%%%%%%%%%%%%%%%%%%%%%%%%%%%%%%%%%%%%%%%%%%%%%%%%%%
\section{Theoretical model}\label{sec2}%{Field equations and their solutions}
%%%%%%%%%%%%%%%%%%%%%%%%%%%%%%%%%%%%%%%%%%%%%%%%%%%%%%%%%%%%%%%%%%%%%%%%%%%%%%%%%
The Einstein field equations for a FRW space-time (with flat spatial section) are given by
\be\label{fe1}
3\frac{{\dot{a}}^{2}}{a^2}=\rho_{m} + \frac{1}{2}{\dot{\phi}}^{2} + V(\phi) = \rho_{m} + \rho_{\phi}
\ee
\be\label{fe2}
2\frac{\ddot{a}}{a}+\frac{{\dot{a}}^{2}}{a^2}=-\frac{1}{2}{\dot{\phi}}^{2} + V(\phi) = - p_{\phi}
\ee
written in natural units such that $8\pi G = c = 1$.\\ It is clear from equations (\ref{fe1}) and (\ref{fe2}) that the energy density $\rho_{\phi}$ and pressure $p_{\phi}$ for the scalar field component are 
\be\label{eqrp1}
\rho_{\phi} = \frac{1}{2}{\dot{\phi}}^{2} + V(\phi) 
\ee
\be\label{eqpp1}
p_{\phi}=\frac{1}{2}{\dot{\phi}}^{2} - V(\phi) 
\ee
Also, the conservation equations for the scalar field and the matter field are 
\be\label{ce1}
{\dot{\rho}}_{\phi} + 3H(\rho_{\phi} + p_{\phi})=0  
\ee
\be\label{ce2}
{\dot{\rho}}_{m} + 3H\rho_{m}=0 
\ee
Equation (\ref{ce2}) on integration yields
\be\label{cem2}
\rho_{m}=\rho_{m0}a^{-3} 
\ee
where $\rho_{m0 }$ denotes the current value of the energy density corresponding to the matter field.\\
%%%%%%%%%%%%%%%%%%%%%%%%%%%%%%%%%%%%%%%%%%%%%%%%%%%%%%%%%%%%%%%
%%%%%%%%%%%%%%%%%%%%%%%%%%%%%%%%%%%%%%%%%%%%%%%%%%%%%%%%%%%%%%%
Also, from equation (\ref{ce1}), one can obtain the {\it equation of state} parameter as
\be\label{wphirho} 
\omega_{\phi}=\frac{p_{\phi}}{\rho_{\phi}}=-1-\frac{a}{3\rho_{\phi}}\frac{d\rho_{\phi}}{da}
\ee
%%%%%%%%%%%%%%%%%%%%%%%%%%%%%%%%%%%%%%%%%%%%%%%%%%%%%%%%%%%%%%%%%%%
%%%%%%%%%%%%%%%%%%%%%%%%%%%%%%%%%%%%%%%%%%%%%%%%%%%%%%%%%%%%%%%%%%%
\par Only three equations amongst (\ref{fe1}), (\ref{fe2}), (\ref{ce1}) and (\ref{ce2}) are independent. The fourth one can be derived from the other three in view of the Bianchi identities. So, we have to solve for four unknown parameters, namely, $H$, $\rho_{m}$, $\phi$ and $V(\phi)$ from three independent equations. Hence, an exact solution is not possible without an additional input. With this freedom, we make an ansatz for the functional form of $\rho_{\phi}$ as,
\be\label{eans1}
\frac{1}{\rho_{\phi}}\frac{d\rho_{\phi}}{da}=-\frac{\lambda a}{(k+a)^{2}}, ~~~~~~k, \lambda ~\rm{~are ~positive ~constants.}
\ee
This immediately yields,
\be\label{eans2}
\rho_{\phi}=\frac{A}{(k+a)^{\lambda}}{\rm exp}{\left[-\frac{k\lambda}{(k+a)}\right]}
\ee
where $A = \rho_{\phi 0} (1+k)^{\lambda} {\rm exp}{\left[\frac{k\lambda}{(1+k)}\right]}$ and $\rho_{\phi 0}$ represents the current value of the scalar field energy density. Of course, the choice made in equation (\ref{eans1}) is quite arbitrary. However, for $k=0$, equation (\ref{eans2}) will provide a simple power law evolution of $\rho_{\phi}$ ($\sim a^{-\lambda}$), which has been considered in many cosmological analysis (Copeland et al. 2006). 
\par From equations (\ref{wphirho}) and (\ref{eans1}), one can immediately obtain the EoS parameter $\omega_{\phi}$ as function of redshift $z$ ($z=\frac{1}{a} -1$) as 
\be\label{wphia}
\omega_{\phi}(z)=-1 + \frac{\lambda}{3{\left[1+k(1+z)\right]}^{2}}
\ee
%%%%%%%%%%%%%%%%%%%%%%%%%%%%%%%%%%%%%%%%%%%%%%%%%%%%%%%%%%%%%%%%%%%%%%%%%%%%%%%%%%%
Infact one can reframe this particular phenomenological model or the phenomenological choice made in (\ref{eans1}) in a different way as well. One can as well make a choice for the equation of state parameter $\omega_{\phi}(z)$ as
\be\label{wphift}
\omega_{\phi}(z)= \omega_0 + \frac{\omega_1}{(\omega_2 + \omega_3 z)^2}
\ee
and for proper choices of $\omega_0$, $\omega_1$, $\omega_2$ and $\omega_3$ one can get back equation (\ref{wphia}).\\
Equation (\ref{wphift}) provides a new form of parametrization for the DE equation of state parameter. It deserves mention that for proper choices of $\lambda$ and $k$ in equation (\ref{wphia}), or equivalently for $\omega_2 = \omega_3 = 1$, equation (\ref{wphift}) takes the form 
\be
\omega_{\phi}(a) = \omega_0 + \omega_1 a^2
\ee
which has been studied extensively in many cosmological DE models (Copeland et al. 2006). However, this representation in terms of the equation of state parameter or energy density of the dark energy sector $\rho_{\phi}(z)$ are interrelated and one can consider any of these approaches to begin with.\\
%%%%%%%%%%%%%%%%%%%%%%%%%%%%%%%%%%%%%%%%%%%%%%%%%%%%%%%%%%%%%%%%%%%%%%%%%%%%%%%%%%%
At present, most of the existing models of dark energy lacks a well motivated physical background which can explain the origin of the late-time cosmic acceleration successfully. So, it is reasonable to consider a phenomenological approach. Cosmologists are looking forward to the DESI (Aghamousa et al. 2016), Euclid (Laureijs et al. 2011) and LSST (Abell et al. 2009) experiments which, when operational, will provide high precision data which will be useful to understand the expansion history of the universe and one will be able to verify the viability of various dark energy models beyond a $\Lambda$CDM model. Until then one can test a cosmological toy model with the available data and check its viability. Motivated by these facts, in this paper, we made the ansatz (\ref{eans1}) to track the expansion dynamics of the universe. The assumption of equation (\ref{eans1}) (or equivalently equation (\ref{eans2}) or (\ref{wphia})) now makes the system of equations closed. In what follows, we shall try to obtain some cosmological solutions for this toy model providing an accelerating universe.\\
%%%%%%%%%%%%%%%%%%%%%%%%%%%%%%%%%%%%%%%%%%%%%%%%%%%%%%%%%%%%%%%%%%%%%%%%%%%%%%%%%%%%%%
\par From equations (\ref{fe1}), (\ref{cem2}) and (\ref{eans2}), the expression for Hubble parameter is obtained as
\be\label{eh}
H^{2}=H^{2}_{0}{\left[ \Omega_{m0}a^{-3} + \frac{\beta \Omega_{\phi 0}}{(k+a)^{\lambda}}{\rm exp}{\left[-\frac{k\lambda}{(k+a)}\right]} \right]}
\ee
where $\beta=(1+k)^{\lambda} {\rm exp}{\left[\frac{k\lambda}{(1+k)}\right]}$ is a constant, $\Omega_{m0}=\frac{\rho_{m0}}{3H^{2}_{0}}$ and $\Omega_{\phi 0}(=\frac{\rho_{\phi 0}}{3H^{2}_{0}})=1-\Omega_{m0}$ represent the current values of the density parameters for the matter and the scalar fields respectively.\\ 
\par The deceleration parameter $q$ is defined as
\be\label{eq}
q=-\frac{\ddot{a}}{aH^{2}}=-{\left(1+\frac{\dot{H}}{H^{2}}\right)}
\ee
where $\dot{H}=\frac{dH}{dt}=aH\frac{dH}{da}$.\\
From equations (\ref{eh}) and (\ref{eq}), we have obtained the expression for $q$ in terms of scale factor $a$ as,
\be\label{eq2}
q(a)=-1 + \frac{\frac{3}{2}\Omega_{m0}a^{-3} + \frac{\lambda \beta \Omega_{\phi 0}a^{2}}{2(k+a)^{\lambda + 2}}{\rm exp}{\left[-\frac{k\lambda}{(k+a)}\right]}}{ \Omega_{m0}a^{-3} + \frac{\beta \Omega_{\phi 0}}{(k+a)^{\lambda}}{\rm exp}{\left[-\frac{k\lambda}{(k+a)}\right]} }
\ee
Now, equation (\ref{eq2}) can be written in terms of redshift $z$ as
\be\label{eq3}
q(z)=-1 + \frac{\frac{3}{2}\Omega_{m0}(1+z)^{3} + \frac{\lambda \beta \Omega_{\phi 0}(1+z)^{\lambda}}{2(1+k(1+z))^{\lambda + 2}}{\rm exp}{\left[-\frac{k\lambda (1+z)}{(1+k(1+z))}\right]}}{ \Omega_{m0}(1+z)^{3} + \frac{\beta \Omega_{\phi 0} (1+z)^{\lambda}}{(1+k(1+z))^{\lambda}}{\rm exp}{\left[-\frac{k\lambda (1+z)}{(1+k(1+z))}\right]} }
\ee
%%%%%%%%%%%%%%%%%%%%%%%%%%%%%%%%%%%%%%%%%%%%%%%%%%%%
For the sake of completeness, we have also obtained the functional behaviour of the density parameters
for the matter field ($\Omega_{m}$) and scalar field ($\Omega_{\phi}$) as,
\bea
\Omega_{m}(z)=\frac{\Omega_{m0}(1+z)^{3}}{\Omega_{m0}(1+z)^{3} + \frac{\beta \Omega_{\phi 0} (1+z)^{\lambda}}{(1+k(1+z))^{\lambda}}{\rm exp}{\left[-\frac{k\lambda (1+z)}{(1+k(1+z))}\right]}}\\
\Omega_{\phi}(z)=\frac{\beta \Omega_{\phi 0}(1+z)^{\lambda} {\rm exp}{\left[-\frac{k\lambda (1+z)}{(1+k(1+z))}\right]}{(1+k(1+z))}^{-\lambda}}{\Omega_{m0}(1+z)^{3} + \frac{\beta \Omega_{\phi 0} (1+z)^{\lambda}}{(1+k(1+z))^{\lambda}}{\rm exp}{\left[-\frac{k\lambda (1+z)}{(1+k(1+z))}\right]}}
\eea
%%%%%%%%%%%%%%%%%%%%%%%%%%%%%%%%%%%%%%%%%%%%%%%%%%%%%%%%%%%%%%%%%%%%%%%%
Now, adding equations (\ref{eqrp1}) and (\ref{eqpp1}), one can obtain
\bea
{\dot{\phi}}^{2}=(1+z)^2H^2{\left(\frac{d\phi}{dz}\right)^2}= (1+\omega_{\phi}(z))\rho_{\phi}(z) \nonumber \\
\Rightarrow \frac{d\phi(z)}{dz}= \pm\frac{\sqrt{\lambda}(1+z)^{-1}}{1+k(1+z)}{\left[1+ \frac{\Omega_{m0}}{\beta \Omega_{\phi 0}}\frac{(1+k(1+z))^\lambda}{(1+z)^{(\lambda -3)}}{\rm exp}{\left[\frac{k\lambda (1+z)}{(1+k(1+z))}\right]}\right]}^{-1/2}
\eea
which on integration gives,
\be\label{eqrcpz1}
\phi(z)=\phi_{0}\pm {\left(\frac{2}{k}\right)}^{\frac{\lambda}{2}} {\lambda^{\frac{(1-\lambda)}{2}}} {\cal F}(z) {\frac{(1+z)^{-1}(1+k(1+z))^{\frac{\lambda}{2}}}{\sqrt{1+{\frac{\Omega_{m0}}{\beta \Omega_{\phi 0}} \frac{(1+k(1+z))^{\lambda}}{(1+z)^{-2}} {\rm exp}{\left[-\frac{k\lambda (1+z)}{(1+k(1+z))}\right]} }}}}
\ee
where, $\phi_{0}$ is an integration constant and ${\cal F}(z)= {\rm exp}{\left[\frac{k\lambda (1+z)}{2(1+k(1+z))}\right]} \Gamma {\left(\frac{\lambda}{2},\frac{k\lambda (1+z)}{2(1+k(1+z))}\right)}$.\\
Similarly, using equations (\ref{eqrp1}) and (\ref{eqpp1}), one can reconstruct the potential for the scalar field as
\be
V(\phi) = \frac{1}{2}\rho_{\phi}(1 - \omega_{\phi})
\ee
which when expressed in terms of redshift parameter $z$ becomes
\be\label{eqrcvz1}
V(z)=V_{0}\frac{(1+z)^{\lambda}}{\left[1+k(1+z)\right]^{\lambda}}{\left[1- \frac{\lambda}{6(1+k(1+z))^2}\right]}{\rm exp}{\left[-\frac{k\lambda (1+z)}{(1+k(1+z))}\right]}
\ee
where, $V_{0}=3H^{2}_{0}\Omega_{\phi 0}\beta$. Therefore, by using equations (\ref{eqrcpz1}) and (\ref{eqrcvz1}), one can arrive at the expression for the potential $V(\phi)$ if the values of $k$ and $\lambda$ are given. In this work, we first obtain constraints on $k$ and $\lambda$ using the observational datasets and from the best-fit values, we then reconstruct the functional form of $V(\phi)$ (see section \ref{result}).\\
%%%%%%%%%%%%%%%%%%%%%%%%%%%%%%%%%%%%%%%%%%%%%%%%%%%%%%%%%%%%%%%%%%%%%%%%%%%%%%%%%%%%%%
\par In order to facilitate the structure formation, an accelerating model of the universe should have a deceleration history in the past as well. So, the deceleration parameter $q$ is an important factor in depicting the evolution history of our universe. For this reason we shall try to analyse the behavior of $q$ for this particular model.
%%%%%%%%%%%%%%%%%%%%%%%%%%%%%%%%%%%%%%%%%%%%%%%%%%%%%%%%%%%%%%%%%%%%%%%%%%%%%%%%%%%%%%%
%%%%%%%%%%%%%%%%%%%%%%%%%%%%%%%%%%%%%%%%%%%%%%%%%%%%%
\section{Data analysis }\label{data}
%%%%%%%%%%%%%%%%%%%%%%%%%%%%%%%%%%%%%%%%%%%%%%%%%
Here we shall fit the present model by using the type Ia supernova (SNIa) dataset and the observational data from Hubble data survey. We present a brief summary of data analysis method for each of the datasets.\\
%%%%%%%%%%%%%%%%%%%%%%%%%%%%%%%%%%%%%%%%%%%%%%%%%%%%%%%%%%%%%%%%%%%%%%%%%%%%%%%%%%%%%
\par For the SNIa dataset, we have used the recently released Union2.1 compilation data (Suzuki et al. 2012) of 580 data points. The corresponding $\chi^2$ function is defined as (Nesseris \& Perivolaropoulos 2005)
\be 
\chi^2_{SN}= A - \frac{B^2}{C}
\ee
with
\bea
A = \sum^{580}_{i=1} \frac{[{\mu}^{obs}(z_{i}) - {\mu}^{th}(z_{i})]^2}{\sigma^2_{\mu}(z_{i})}\\
B= \sum^{580}_{i=1} \frac{[{\mu}^{obs}(z_i) - {\mu}^{th}(z_{i})]}{\sigma^2_{\mu}(z_{i})}
\eea
and
\be 
C= \sum^{580}_{i=1} \frac{1}{\sigma^2_{\mu}(z_{i})}
\ee 
where $\mu^{obs}$ is the observed distance modulus at a particular redshift, $\mu^{th}$ is the corresponding theoretical counterpart and $\sigma_{\mu}$ is the error. \\
%%%%%%%%%%%%%%%%%%%%%%%%%%%%%%%%%%%%%%%%%%%%%%%%%%%%%%%%%%%%%%%%%%%%%%%%%%%%%%%%%%%%%%% 
\par Next, we have continued the analysis with the 29 data points obtained in Hubble parameter measurements (Simon et al. 2005; Stern et al. 2010; Blake et al. 2012; Moresco et al. 2012; Chuang \& Wang 2013;  Samushia et al. 2013; Zhang et al. 2014; Delubac et al. 2015; Ding et al. 2015) in the range $0.07\le z\le 2.34$ (Mamon \& Das 2015). The corresponding $\chi^2$ function is given by
\be
\chi^2_{H} = \sum^{29}_{i=1}\frac{[{h}^{obs}(z_{i}) - {h}^{th}(z_{i})]^2}{\sigma^2(z_{i})}
\ee
In the above equation,  $h^{obs}$ and $h^{th}$ are the observed and theoretical values of the Hubble parameter respectively. Also, $\sigma$ represents the error in Hubble parameter measurements and $h(z)= \frac{H(z)}{H_{0}}$.\\ 
Now the total $\chi^2$ for the (SNIa+Hubble) dataset is defined as 
\be 
\chi^2_{total}= \chi^2_{SN} + \chi^2_{H}
\ee
%%%%%%%%%%%%%%%%%%%%%%%%%%%%%%%%%%%%%%%%%%%%%%%%%%%%%%%%%%%%%%%%%%%%%%%%%%%%%%%%%%%%%%%%
One can now minimize these $\chi^{2}$ functions (i.e., $\chi^2_{SN}$, $\chi^2_{H}$ and $\chi^2_{total}$) in respect of the model parameters and compute the  estimated values and their errors.
%%%%%%%%%%%%%%%%%%%%%%%%%%%%%%%%%%%%%%%%%%%%%%%%%%%%
\section{Results}\label{result}
\begin{figure}[ht]
\begin{center}
\includegraphics[width=0.32\columnwidth]{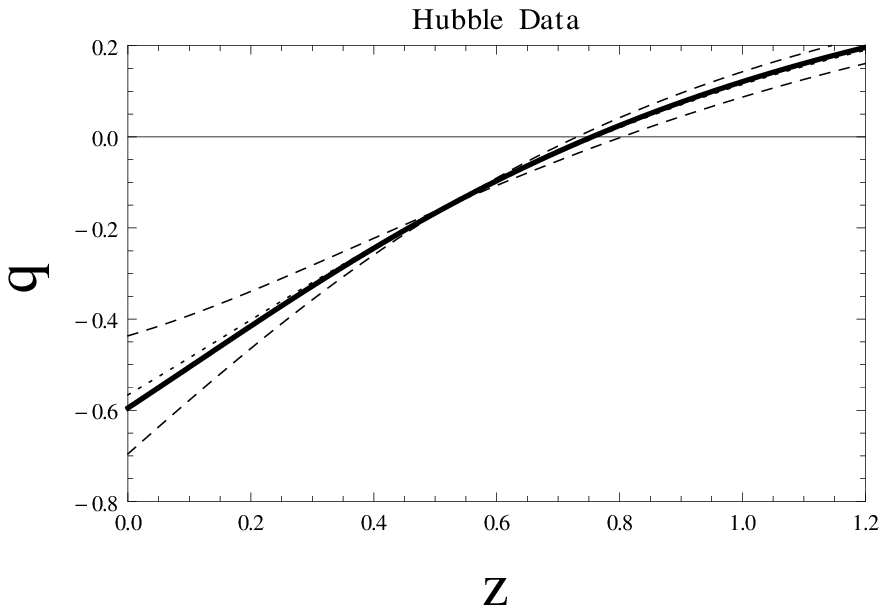}
\includegraphics[width=0.32\columnwidth]{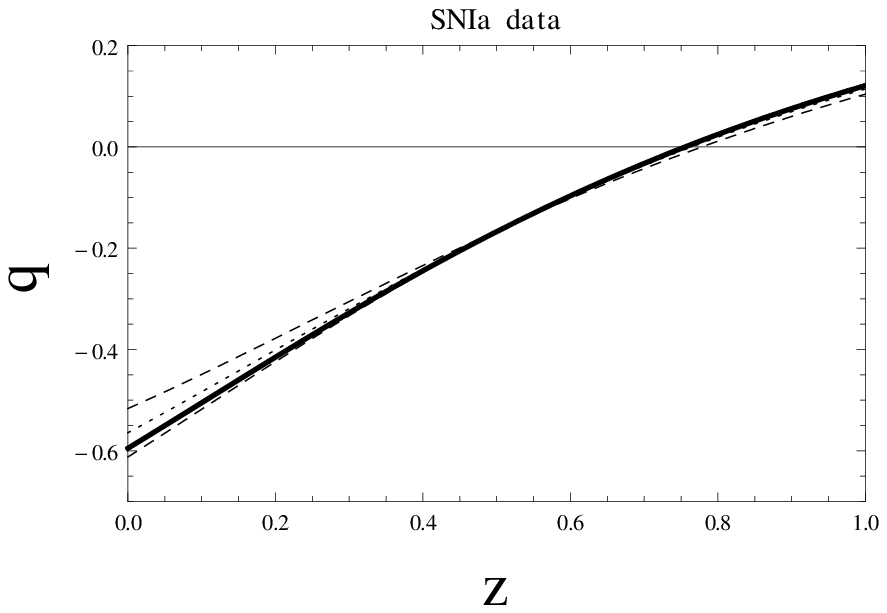}
\includegraphics[width=0.32\columnwidth]{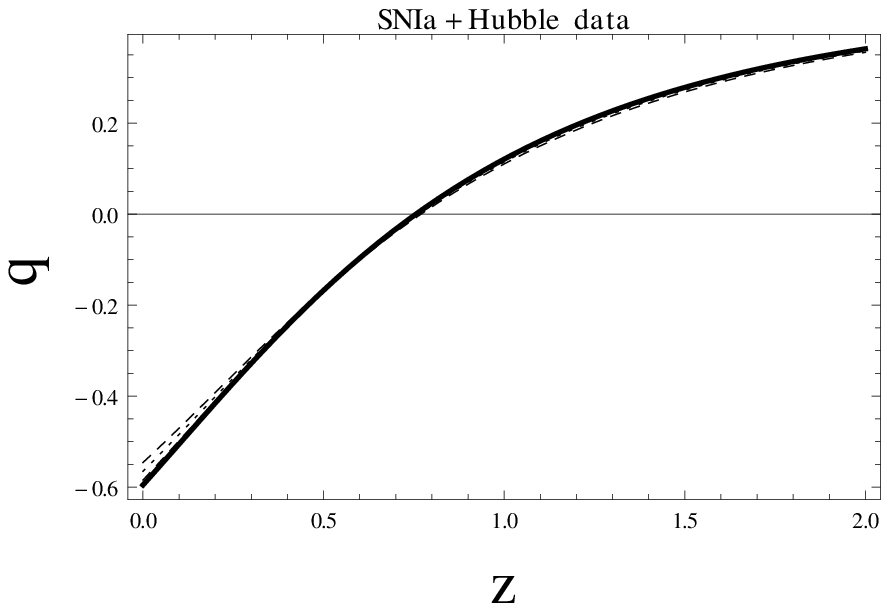}
\caption{\em The reconstructed $q(z)$ for different observational datasets are shown. For each panel, the central dotted line and the dashed lines represent the best-fit curve with $1\sigma$ errors. Also, in each panel, the thick line indicates a $\Lambda$CDM universe (with $\Omega_{m0}=0.27$ and $\Omega_{\Lambda 0}=0.73$). This is for $\Omega_{m0} = 0.27$.}
\label{figqz}
\end{center}
\end{figure}

%%%%%%%%%%%%%%%%%%%%%%%%%%%%%%%%%%%%%%%%%%%%%%%%%%%%%%%%%%%%%%%%%%%%%%%%
Following the data analysis method mentioned above, in this section, limits on the values of $k$ and $\lambda$ are obtained for the Hubble, SNIa and Hubble+SNIa datasets which are displayed in the table \ref{tab:fntable1} alongwith the $1\sigma$ errors. 
%%%%%%%%%%%%%%%%%%%%%%%%%%%%%%%%%%%%%%%%%%%%%%%%%%%%%%%
\begin{table}[ht]
\begin{center}
%\begin{fntable}[0.8\columnwidth]
\begin{tabular}{|c|c|c|c|}
\hline
Datasets & $k$ & $\lambda$ & $\chi^{2}_{m}$ (minimum value of $\chi^{2}$) \\ \hline
Hubble & $4.96 \pm 0.40$ & $2.82 \pm 0.23$ & $28.59$ \\ \hline
SNIa & $4.97 \pm 0.22$ & $2.99 \pm 0.19$ & $562.27$\\ \hline
SNIa+Hubble &  $4.93 \pm 0.10$ & $2.94 \pm 0.12$ & $573.84$\\ 
\hline
\end{tabular}
%\end{fntable}
\caption{\em Best fit values for $k$ and $\lambda$ for the Hubble and SNIa datasets with $\Omega_{m0} = 0.27$. Here, $\chi^{2}_{m}$ represents the minimum value of $\chi^{2}$.}
\label{tab:fntable1} 
\end{center} 
\end{table}
%%%%%%%%%%%%%%%%%%%%%%%%%%%%%%%%%%%%%%%%%%%%%%%%%%%%%%%%%%
It has been found that the joint analysis of the SNIa+Hubble dataset put a tighter constraint as compared to the constraints obtained from SNIa or Hubble dataset alone. Using these values, the deceleration parameter $q(z)$ has been reconstructed for different datasets which are shown in figure \ref{figqz}.
%%%%%%%%%%%%%%%%%%%%%%%%%%%%%%%%%%%%%%%%%%%%%%%%
%%%%%%%%%%%%%%%%%%%%%%%%%%%%%%%%%%%%%%%%%%%%%%%%%%%%
%%%%%%%%%%%%%%%%%%%%%%%%%%%%%%%%%%%%%%%%%%%%%%%%
\begin{figure}[ht]
\begin{center}
\includegraphics[width=0.35\columnwidth]{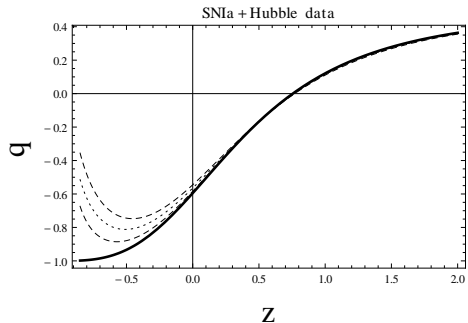}
\caption{\em Future evolution of $q(z)$ for this model shown by dotted line along the $1\sigma$ contour (dashed lines). This plot corresponds to values of ($k$, $\lambda$) obtained for the SNIa+Hubble dataset with $\Omega_{m0} = 0.27$. The thick line as usual indicates the behaviour of $q$ for $\Lambda$CDM model.}% The central dotted line represents the best-fit curve and the dashed contour represents $1\sigma$ confidence level. Also, the thick line represents the evolution of $q$ in a flat $\Lambda$CDM.}
\label{figqzf}
\end{center}
\end{figure} 
%%%%%%%
From figure \ref{figqz}, we have found that $q(z)$ enters into a negative value regime in the recent past at a redshift $z_{t}$. The best-fit values of $q(z)$ at present (say, $q_{0}=q(z=0)$) and the redshift $z_{t}$ at which transition in $q$ occurs alongwith $1\sigma$ errors for different datasets are listed in table \ref{tab:fntable2}.  
%%%%%%%%%%%%%%%%%%%%%%%%%%%%%%%%%%%%%%%%%%%%%%%%%%%%%%%%%%%%%%%%%%%%%%
\begin{table}[ht]
\begin{center}
%\begin{fntable}[0.8\columnwidth]
\begin{tabular}{|c|c|c|}
\hline
Datasets & $q_0$ & $z_t$ \\ \hline
Hubble & $q_{0}=-0.57 \pm 0.13$ & $z_{t}=0.75 \pm 0.04$\\ \hline
SNIa & $q_{0}=-0.56 \pm 0.05$ & $z_{t}=0.76 \pm 0.02$\\ \hline
SNIa+Hubble & $q_{0}=-0.56 \pm 0.02$  & $z_{t}=0.76 \pm 0.01$\\ 
\hline
\end{tabular}
%\end{fntable}
\caption{\em Best fit values of $q_0$ and $z_t$ (within $1\sigma$ errors) for different datasets.}
\label{tab:fntable2} 
\end{center} 
\end{table}
%%%%%%%%%%%%%%%%%%%%%%%%%%%%%%%%%%%%%%%%%%%%%%%%%%%%%%%%%%%%%%%%%%%%%

This results are almost consistent with the values known for the flat $\Lambda$CDM model ($q_{0}=-0.59$, $z_{t}=0.75$) with $\Omega_{m0}=0.27$ and $\Omega_{\Lambda 0}=0.73$. It deserves mention that our results also match with that obtained in literature [for details, one can look at Refs. (Turner \& Riess 2002; Riess 2004; Chuna 2009; Mamon \& Das 2016) and the references therein]. 
%%%%%%%%%%%%%%%%%%%%%%%%%%%%%%%%%%%%%%%%%%%%%%%%%%%%%%%%%%
\begin{figure}[ht]
\begin{center}
\includegraphics[width=0.32\columnwidth]{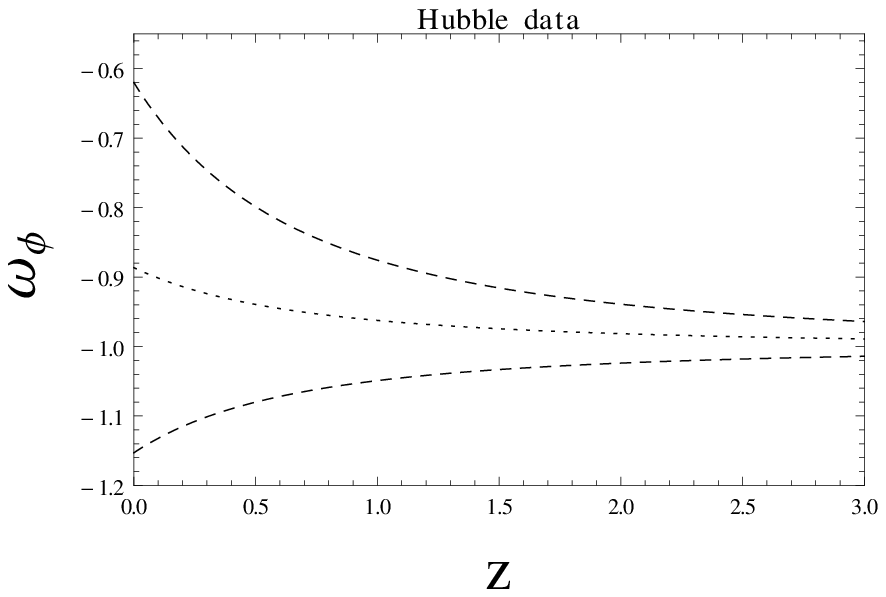}
\includegraphics[width=0.32\columnwidth]{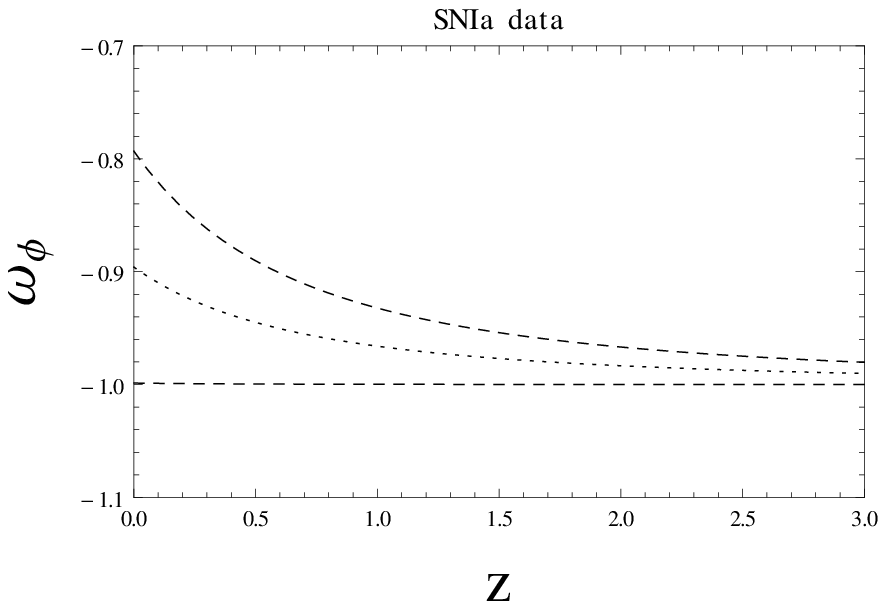}
\includegraphics[width=0.32\columnwidth]{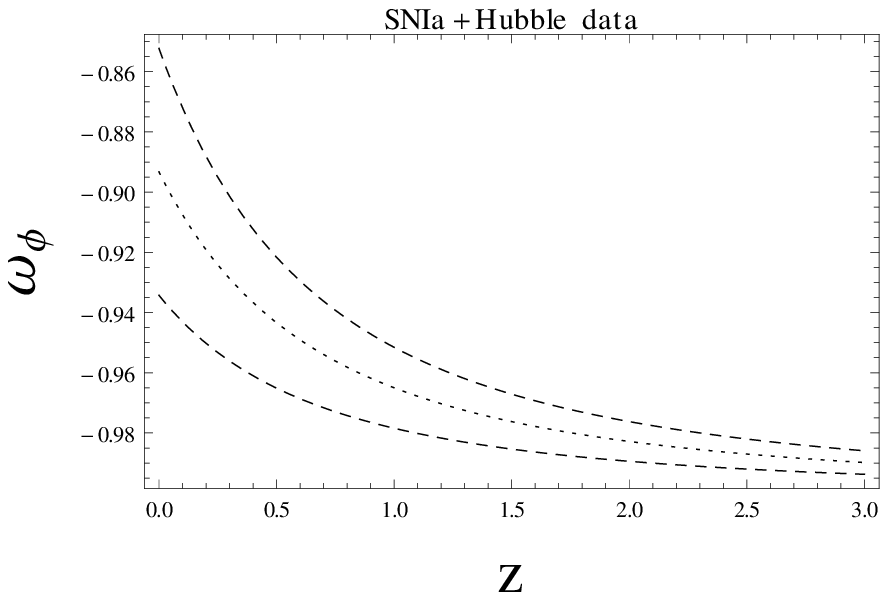}
\caption{\em The reconstructed EoS parameter $\omega_{\phi}(z)$ for this model using various observational datasets, as indicated in each panel. The central dotted line represents the best-fit curve and the dashed lines represent $1\sigma$ contour. All the plots are for $\Omega_{m0} = 0.27$. }
\label{figomegaz}
\end{center}
\end{figure}
%%%%%%%%%%%%%%%%%%%%%%%%%%%%%%%%%%%%%%%%%%%%%%%%%%%%
Figure \ref{figqzf} shows the future evolution of $q(z)$. It is evident from figure \ref{figqzf} that the present model does not show any indication of slowing down of the present cosmic acceleration in near future as suggested in Refs. (Shafieloo et al. 2009; Magana et al. 2014) for various dark energy parametrizations. In far future (near $z \rightarrow -1$), however there is evidence that the rate of expansion varies but the universe continues to accelerate forever in the present toy model. Hence, we need more  robust observational datasets and more effective analysis methods to have consensus on whether the cosmic acceleration is speeding up or not.\\
%%%%%%%%%%%%%%%%%%%%%%%%%%%%%%%%%%%%%%%%%%%%%%%%%%%%%%%%%%%%%%%%
The reconstructed evolution dynamics of $\omega_{\phi}(z)$  is shown in
figure \ref{figomegaz} for different datasets. The values of $\omega_{\phi}(z)$ at present (i.e., $\omega_{\phi}(z=0)$) with $1\sigma$ errors for the Hubble, SNIa and SNIa+ Hubble datasets are obtained as $-0.88\pm 0.26$, $-0.89\pm 0.10$ and $-0.89\pm 0.04$ respectively. 
%%%%%%%%%%%%%%%%%%%%%%%%%%%%%%%%%%%%%%%%%%%%%%%%%%%%%%%%%%
\begin{figure}[ht]
\begin{center}
\includegraphics[width=0.4\columnwidth]{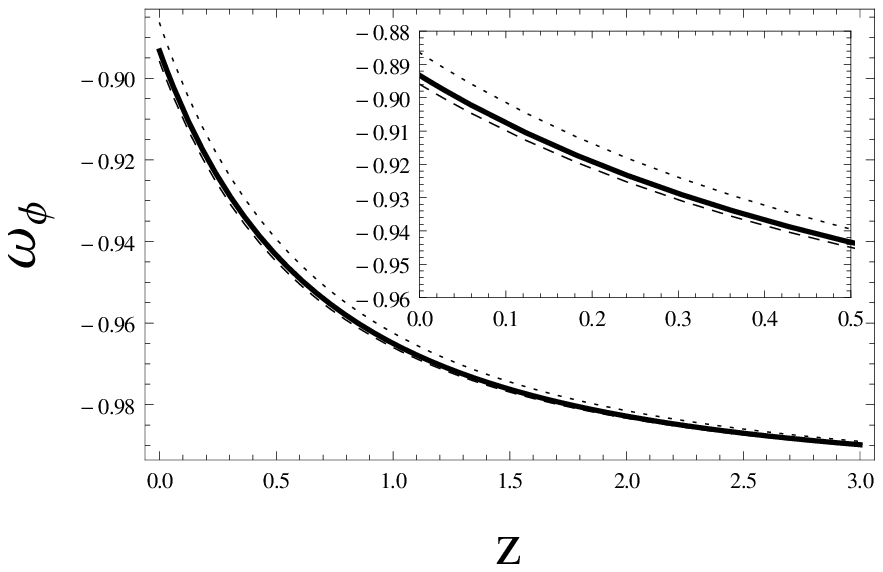}
\includegraphics[width=0.4\columnwidth]{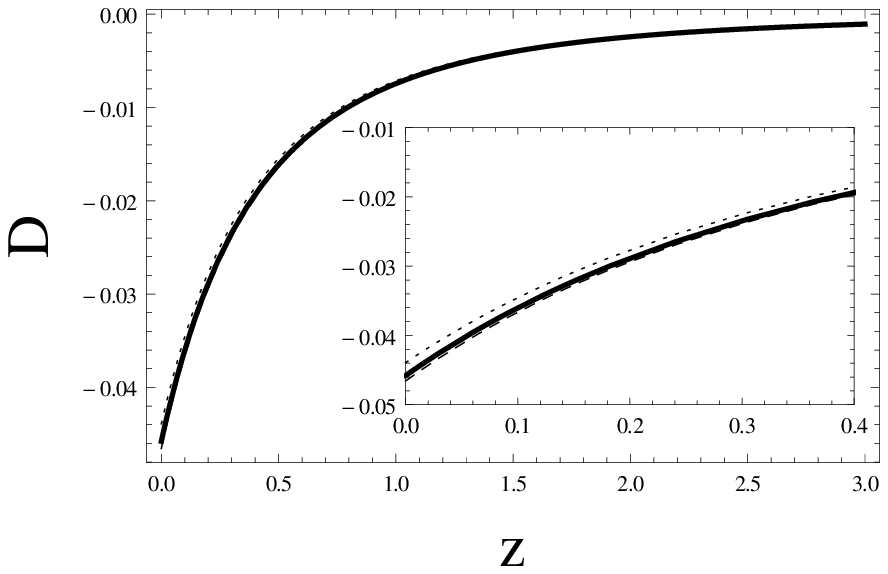}
\caption{\em The left panel shows the plot of the reconstructed EoS parameter $\omega_{\phi}(z)$ using the best-fit values of $k$ and $\lambda$ and $\Omega_{m0} = 0.27$. The right panel shows the plot of $D{\left(=\frac{d\omega_{\phi}}{dz}\right)}$ against $z$. In both plots, the dotted, dashed and thick lines show the evolution of the corresponding parameter for the Hubble, SNIa and SNIa+Hubble datasets respectively. }
\label{figomegabf}
\end{center}
\end{figure}
%%%%%%%%%%%%%%%%%%%%%%%%%%%%%%%%%%%%%%%%%%%%%%%%%%%% 
In left panel of figure \ref{figomegabf}, we have shown the behavior of  $\omega_{\phi}(z)$  for the values of $k$ and $\lambda$ obtained in table \ref{tab:fntable1} for each dataset. 
We have also plotted the rate of change of $\omega_{\phi}$ against $z$ ${\left(D =\frac{d\omega_{\phi}}{dz}\right)}$ in figure \ref{figomegabf}. It shows that the magnitude of $\frac{d\omega_{\phi}}{dz}$ is negative and remains almost constant at high redshifts, but the magnitude of $\frac{d\omega_{\phi}}{dz}$ is decreasing at low redshifts for each dataset. Figure \ref{figomegaz} and figure \ref{figomegabf} indicate that at high redshifts the present model does not have any significant deviation from $\Lambda$CDM model, but with evolution (as $z \rightarrow 0$), the deviation from $\Lambda$CDM becomes prominent. This dynamical nature of DE component can be effective in determining the late time evolution of the universe and thus may provide answer to the {\it coincidence problem} in cosmology.
%%%%%%%%%%%%%%%%%%%%%%%%%%%%%%%%%%%%%%%%%%%%%%%%%%%%%%%%%%%%%%%%%
\par
\begin{figure}[ht]
\begin{center}
\includegraphics[width=0.3\columnwidth]{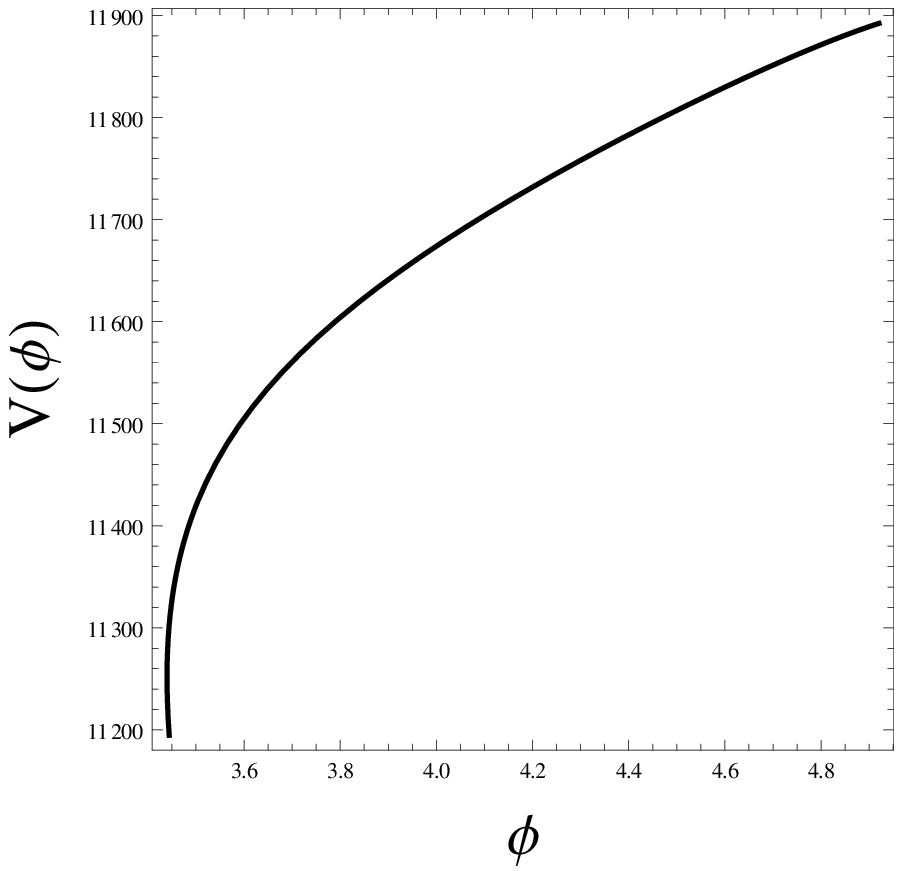}
\includegraphics[width=0.3\columnwidth]{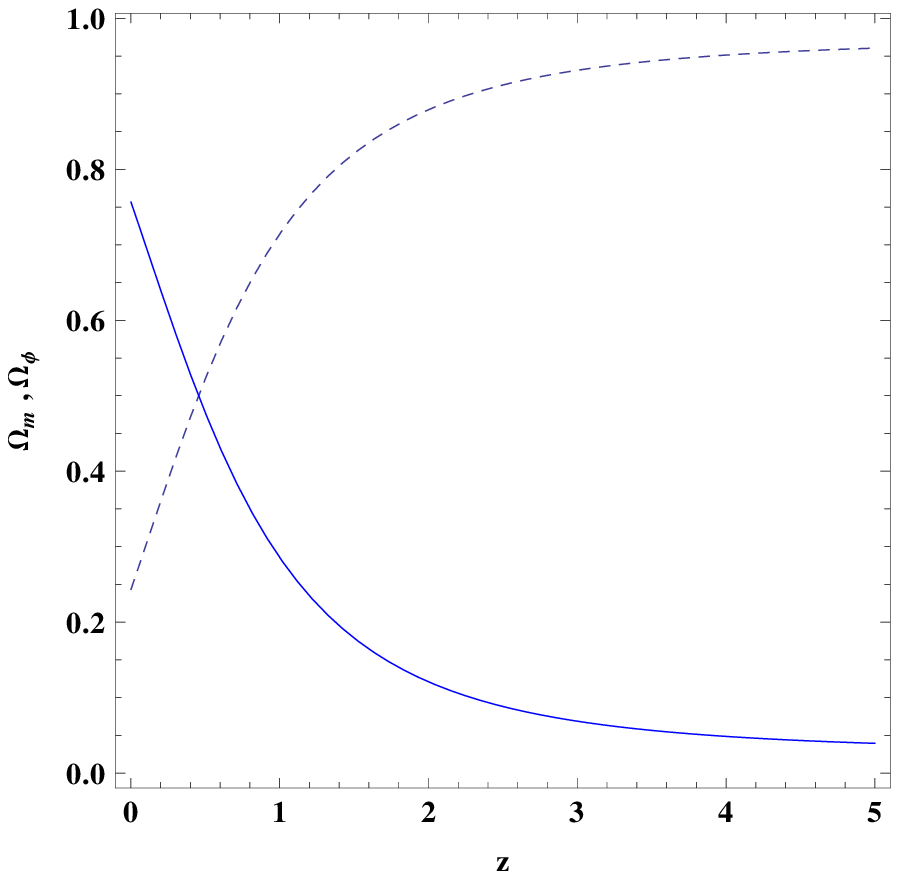}
\caption{\em The left panel shows the reconstructed potential $V(\phi)$ with the ($k$, $\lambda$) values arising from the SNIa+Hubble dataset. In this plot, we have chosen $\Omega_{\phi 0}=0.73$, $H_{0}=72~km/s/Mpc$ and $\phi_{0} =5$. The right panel shows the plot of $\Omega_{m}$ (dashed curve) and $\Omega_{\phi}$ (solid curve) for $k=4.9$, $\lambda=2.9$ and $\Omega_{m0} = 0.27$.}
\label{figomega}
\end{center}
\end{figure}

 For the sake of completeness, we have also solved equations (\ref{eqrcpz1}) and (\ref{eqrcvz1}) numerically and plotted the potential $V(\phi)$ for $k=4.93$, $\lambda=2.94$, $\Omega_{\phi 0}=0.73$, $H_{0}=72~km/s/Mpc$ and $\phi_{0} =5$ in left panel of figure \ref{figomega}. From this figure, we have found that the potential $V(\phi)$ increases with $\phi$. The reason behind this seems to be the choice of $\rho_{\phi}$ as given in equation (\ref{eans1}). For this toy model, $V(\phi)$ can be obtained as
\be\label{vphieq}
V(\phi)\approx A {\rm exp}(\alpha_{1}\phi) + B {\rm exp}(\alpha_{2}\phi)
\ee
where $A=1.07 \times 10^{4}$, $\alpha_{1}=0.02$, $B=-4.21 \times 10^{16}$ and $\alpha_{2}=-9.50$. Recently, this type of potentials have already been discussed by several authors while explaining the late-time cosmic acceleration (Barreiro et al. 2000; Rubano \&  Sudellaro 2001; Sen \& Sethi 2002). We have also checked that the nature of the $V(\phi)$ curve is hardly affected by a small change in the allowed values of $k$, $\lambda$ within $1\sigma$ confidence limit and other choices of $\phi_{0}$.\\  
The variation of density parameters $\Omega_{m}(z)$ and $\Omega_{\phi}(z)$ are also shown in the right panel of figure \ref{figomega}. This plot also indicates that the universe has evolved to a dark energy dominated era in the recent past, which is in accordance with observational results. 
%%%%%%%%%%%%%%%%%%%%%%%%%%%%%%%%%%%%%%%%%%%%%%%%%%%%%%%%%%%%%%%%%%%%%%%%%%%%%%%%%
%%%%%%%%%%%%%%%%%%%%%%%%%%%%%%%%%%%%%%%%%%%%%%%%%%%%%%%%%%%%%%%%%%%%%%%%%%%%%%%%%%%
%%%%%%%%%%%%%%%%%%%%%%%%%%%%%%%%%%%%%%%%%%%%%%%%%%%%%%%%%%
%%%%%%%%%%%%%%%%%%%%%%%%%%%%%%%%%%%%%%%%%%%%%%%%%%%%%%%%%%%%%%%
\section{Growth of perturbations}\label{perturbation}
%%%%%%%%%%%%%%%%%%%%%%%%%%%%%%%%%%%%%%%%%%%%%%%%%%%%%%%
%%%%%%%%%%%%%%%%%%%%%%%%%%%%%%%%%%%%%%%%%%%%%%%%%%%%%%%%%%%%%%%%%%%%%%%%%%%%%%%%
We are also interested to look into the effect of this particular DE sector on the evolution of matter over-densities. It is expected that the growth of matter perturbations will be effected in presence of a DE sector. As DE sector provides a replusive gravity effect, it will result in the slowing down of the growth of matter sector. However, for different DE models the effect will be different depending upon the nature of the DE equation of state parameter. In this section we want to study the rate by which the evolution of matter densities gets effected for this particular form of DE density.  To study this, we consider the following system of linearized Einstein equations (Jaber 2017):
\bea
a^2 \delta_m''(a) + a\frac{3}{2}\left[1-\omega_{\phi}(a)\Omega_{\phi}(a)\right]\delta_m'(a) - \frac{3}{2} \left[ \Omega_m(a) \delta_m(a) + \Omega_{\phi}(a) \delta_{\phi}(a) \right] = 0  \label{eq:Mcoupled} \\ 
a^2 \delta_{\phi}''(a) + a\frac{3}{2}\left[1-\omega_{\phi}(a)\Omega_{\phi}(a)\right]\delta_{\phi}'(a) + \left(\frac{c_s^2 \kappa^2}{a^2 H^2(a)} - \frac{3}{2} \Omega_{\phi}(a)\right)\delta_{\phi}(a) - \frac{3}{2} \Omega_{m}(a) \delta_m(a) =  0, \label{eq:DEcoupled}
\eea
where $\delta_m\equiv\frac{\delta\rho_m}{\rho_m}$ and $\delta_{\phi}\equiv\frac{\delta\rho_{\phi}}{\rho_{\phi}}$ represent the matter and DE density contrasts, respectively. A prime indicates variation with respect to $a$ and $\kappa$ is the Fourier wave number. Also, the term $c^2_{s}$ in equation (\ref{eq:DEcoupled}) represents the speed of sound for the DE sector. One can split it as the sum of an adiabatic and an effective (non-adiabatic) contribution, namely $c^2_{ad}$ and $c^2_{eff}$ respectively, given by:
\be
c^2_{s}=\frac{\delta p_{\phi}}{\delta \rho_{\phi}}=c^2_{ad}+c^2_{eff}
\ee
where $c^2_{ad}=\omega_{\phi} -\frac{1}{3}\frac{\dot{\omega}_{\phi}}{H (1+\omega_{\phi})} = \omega_{\phi} (a)-\frac{1}{3}\frac{a{\omega}^{\prime}_{\phi}(a)}{(1+\omega_{\phi}(a))}$.  Following (Jaber 2017), in this work, we have modelled $c^2_{eff}$ as a constant which can take values $c^2_{eff}=0,~ \frac{1}{3}~{\rm or}~1$.
\par To solve these system of equations, we need initial conditions for $\delta_m$ and $\delta_{\phi}$. For our case, we set our initial conditions at matter dominant era when the DE contribution was very small and the modes are well inside the horizon. We choose  $\delta_m (a_{ini}) = 10^{-5}$ at $\kappa=0.01 Mpc^{-1}$, which corresponds to the value when the $\kappa$-mode enters the horizon. For the scalar field perturbation, the contribution from DE sector is considered to be negligible initially and is set at $\delta_{\phi} (a_{ini}) = 10^{-8}$. 
\begin{figure}[ht]
\begin{center}
\includegraphics[width=0.35\columnwidth]{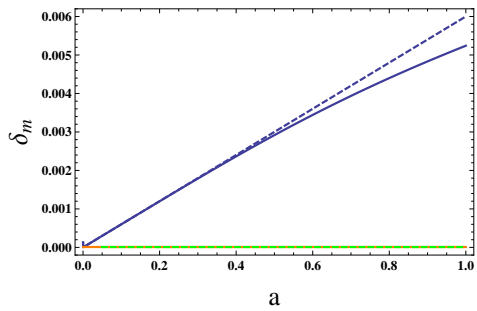}
\caption{\em Growth of matter overdensities $\delta_m (a)$ for $c^2_{s} = 1$. The solid line represents the growth of perturbations for the present DE model whereas the dashed line represents the growth rate in absence of DE. The orange and green lines represent $\delta_m (a)$ for $\Lambda$CDM and CPL models respectively. In this plot, we have chosen $H_{0}=72~km/s/Mpc$ and the values of the model parameters $k$ and $\lambda$ have been taken from joint analysis of SNIa + Hubble dataset as listed in table \ref{tab:fntable1}.}
\label{figdm}
\end{center}
\end{figure}
\par With the initial conditions mentioned above, the system of equations are solved numerically for different values of $c_{eff}^2$. We have displayed the results in figure \ref{figdm} for $c_{eff}^2 = 1$. However, it has been found that the different values of $c_{eff}^2$ only reduces the growth of matter overdensities slightly keeping the shape the same. In figure \ref{figdm} the solid line represents the growth of matter perturbations for the present DE model which is slower compared to the growth rate when there is no DE component in the universe (shown by dotted line in figure \ref{figdm}). We have also compared the growth rate for our model with that for a $\Lambda$CDM model ($\omega_{\Lambda CDM} = -1$)and CPL model ($\omega_{CPL} = \omega_{0} + \frac{\omega_1 z}{(1+z)}$) (Chevallier \& Polarski 2001; Linder 2003) (shown by orange and green lines respectively).  For the CPL model, the values of $\omega_0$ and $\omega_1$ has been taken as $\omega_0 = -1.17$ and $\omega_1 = 0.35$ (Qi et al. 2016). It is evident that with evolution (increasing $a$), the effect of the present DE sector on the growth of matter overdensities is larger as compared to a $\Lambda$CDM or a CPL model.  
\begin{figure}[ht]
\begin{center}
\includegraphics[width=0.35\columnwidth]{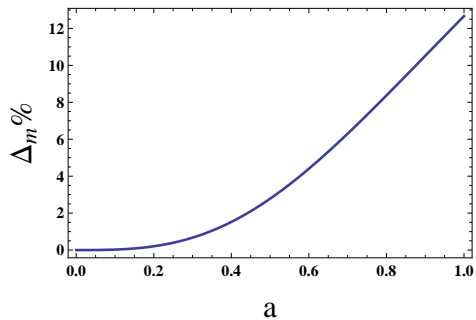}
\caption{\em Percentage decrease in $\delta_m$ as function of scale factor compared to a no DE model}
\label{figpercentage}
\end{center}
\end{figure}
\par In figure \ref{figpercentage} we have plotted the percentage deviation in the growth rate for the present model compared to a no DE model. We have actually plotted the percentage decrease in the growth rate given by $\Delta_m = \frac{\delta_m - \delta_{m(no DE)}}{\delta_{m(no DE)}}$. The higher the percentage decrease, the slower is the growth rate. It is evident from the figure that the growth rate becomes slower with the evolution and at later times when the DE component dominates the evolution, the growth rate is suppressed  by around $12\%$. 
%%%%%%%%%%%%%%%%%%%%%%%%%%%%%%%%%%%%%%%%%%%%%%%%%%%%%%%%%%%%%%%
\section{Conclusion}\label{conclusion}
%%%%%%%%%%%%%%%%%%%%%%%%%%%%%%%%%%%%%%%%%%%%%%%%%%%%%%%
%%%%%%%%%%%%%%%%%%%%%%%%%%%%%%%%%%%%%%%%%%%%%%%%%%%%%%%%%%%%%%%%%%%%%%%%%%%%%%%%
To summarize, in this paper, we have tried to show that a canonical scalar field model can provide an early decelerated expansion followed by an accelerated expansion at late times. For this purpose, we have chosen one specific ansatz for $\rho_{\phi}$ to characterize the properties of DE. Then with this input, we have obtained exact analytical solutions for various cosmological parameters. Using the SNIa, Hubble and SNIa+Hubble datasets, we have reconstructed the deceleration parameter $q(z)$ and the EoS parameter $\omega_{\phi}(z)$ of this model. Results show that the evolution of $q(z)$ does not provide any signal of cosmic deceleration in future. The reconstructed values of $q_{0}$, $z_{t}$ and  $\omega_{\phi}(z=0)$ have been calculated and it has been found that the results obtained do not deviate much from the standard $\Lambda$CDM model. Furthermore, the potential $V(\phi)$ has been found numerically for some specific choices of model parameters and the potential is found to be a combination of two exponentials in $\phi$ (see equation (\ref{vphieq})). As already discussed, this type of potentials have earlier been considered by several authors for quintessence fields. Hence, this work shows again the importance of double exponential potential for a quintessence field. Finally, we would like to mention that the observational datasets suffer from systematic errors and the reconstructed results might vary for other datasets. So, one can hope that the next generation observational datasets will improve the constraints on these model parameters considerably. \\ 
\par From the perturbative analysis it has been found that the dynamical evolution of the DE sector or the corresponding EoS parameter $\omega_{\phi}(z)$ got imprinted in the growth rate of the matter sector and this effect is much prominent at later times for the present DE model as compared to a $\Lambda$CDM model or a CPL model.  
\\
However, as nothing much is known about the DE sector and a wide variety of possibilities are open, various effective cosmological toy models can be considered for different functional forms for $\rho_{\phi}$, which may show even better agreement to the observational results. So one effective way to check the viability of a DE model may be to look at the imprints of these models on the growth rate of matter perturbations and compare it with available experimental measurements. \\
\\
%%%%%%%%%%%%%%%%%%%%%%%%%%%%%%%%%%%%%%%%%%%%%%%%%%%%%%%%%%%%%%%%%%%%%
{\bf Acknowledgements:} SD and MB acknowledge the financial support from SERB, DST, Government of India through the
project EMR/2016/007162. SD would also like to acknowledge IUCAA, Pune for providing support
through associateship programme. AAM acknowledges the financial support from SERB, Government of India through National Post-Doctoral Fellowship Scheme (File No: PDF/2017/000308) and the Department of Physics, Visva-Bharati where a part of the work was completed.
%%%%%%%%%%%%%%%%%%%%%%%%%%%%%%%%%%%%%%%%%%%%%%%%%%%%%%%%%%%%%%%
\section{References}
Abell P. A. et al. 2009, LSST Science Collaboration, LSST Science Book, Version 2.0\\
Aghamousa A. et al. 2016, The DESI Experiment Part I : Science, Targeting and Survey Design\\
Barreiro T., Copeland E. J. \& Nunes N. J. 2000, Phys. Rev. D, 61, 127301 \\
Bartolo N. et al. 2004, Phys. Rev. D {\bf 70}, 043532 \\
Blake C. et al. 2012, MNRAS, {\bf 425}, 405 \\
Chevallier M. \& Polarski D. 2001, Int. J. Mod. Phys. D 10, 213\\
Choudhury T. R. \& Padmanabhan T. 2005, Astron. Astrophys., {\bf 429}, 807\\
Chuang C. H. \& Wang Y. 2013, MNRAS, {\bf 435}, 255 \\
Chuna J.V. 2009, Phys. Rev. D, {\bf 79}, 047301 \\
Copeland E. J., Sami M. \& Tsujikawa S. 2006, Int. J. Mod. Phys. D {\bf 15}, 1753 \\
Delubac T. et al. 2015, A$\&$A, {\bf 574}, A59 \\
Ding X. et al. 2015, ApJ, {\bf 803}, L22 \\
Eisenstein D. J. et al. 2005, Astrophys. J., {\bf 633}, 560 \\
%%%%%%%%%%%%%%%%%%%%%%%%%%%%%%%%%%%%%%
Jaber M. \& Macorra A. de la 2017, preprint (astro-ph.CO/1708.08529)\\
Jassal H. K. 2010, Phys. Rev. D {\bf 81}, 083513 \\
Laureijs R. et al. 2011, Euclid Definition Study Report\\
Linder E.V. 2003, Phys. Rev. Lett. 90, 091301\\
%%%%%%%%%%%%%%%%%%%%%%%%%%%%%%%%%%%%%%%%%%%%%%%%%%%%%%%%%%%%%%%%%%%%%%%%
Magana J. et al. 2014, JCAP, {\bf 017}, 10 \\
Mamon A. A. \& Das S. 2015, Eur. Phys. J. C, {\bf 75}, 244 \\
Mamon A. A. \& Das S. 2016, Int. J. Mod. Phys. D, {\bf 25}, 1650032 \\
Martin J. 2008, Mod. Phys. Lett. A, {\bf 23}, 1252 \\
Moresco M. et al. 2012, JCAP, {\bf 08}, 006 \\
Nesseris S. \& Perivolaropoulos L. 2005, Phys. Rev. D, {\bf 72}, 123519\\ 
%%%%%%%%%%%%%%%%%%%%%%%%%%%%%%%%%%%%%%%%%%%%%%%%%%%%%%%%%%%%%%%%%%%%%%%%
Padmanabhan T. \&  Choudhury T. R. 2003, Mon. Not. R. Astron. Soc., {\bf 344}, 823 \\
Perlmutter S. et al. 1999, Astron. J. {\bf 517}, 565\\
Qi Jing - Zhao et al. 2016, preprint (gr-qc/1606.00168)\\
Riess A. G. et al. 1998, Astron. J. {\bf 116}, 1009\\
Riess A. G. et al. 2004, Astrophys. J., {\bf 607}, 665 \\
Riess A. G. 2001, Astrophys. J., {\bf 560}, 49 \\
Rubano C. \&  Sudellaro P. 2001, preprint (astro-ph/0103335)\\
Sahni V. \& Starobinsky A. A. 2000, Int. J. Mod. Phys. D {\bf 9}, 373\\
Sahni V. 2004, Lect. Notes Phys. {\bf 653}, 141 \\
Samushia L. et al. 2013, MNRAS, {\bf 429}, 1514 \\
Sen A. A. \& Sethi S. 2002, Phys. Lett. B {\bf 532}, 159 \\
Shafieloo A., Sahni V. \& Starobinsky A. A. 2009, Phys. Rev. D, {\bf 80}, 101301\\
Simon J. et al. 2005, Phys. Rev. D, {\bf 71}, 123001 \\
Spergel D. N. et al. 2007, Astrophys. J. Suppl., {\bf 170}, 377  \\
Steinhardt P. J. et al. 1999, Phys. Rev. Lett., {\bf 59}, 123504  \\
Stern D. et al. 2010, JCAP, {\bf 02}, 008 \\
Suzuki N. et al. 2012, Astrophy. J., {\bf 746}, 85 \\ 
Turner M. S. \& Riess A. G. 2002, Astrophys. J., {\bf 569}, 18 \\
%%%%%%%%%%%%%%%%%%%%%%%%%%%%%%%%%%%%%%%%%%%%%%%%%%%%%%%%%%%%%%%%%%%%%%%%
Unnikrishnan S. 2008a, Phys. Rev. D {\bf 78}, 063007\\
Unnikrishnan S., Jassal H. K. \& Seshadri T. R. 2008b, Phys. Rev. D {\bf 78}, 123504 \\
Weinberg S. 1989, Rev. Mod. Phys. {\bf 61}, 1 \\
Weller J. \& Lewis A. M. 2003, Mon. Not. Roy. Astron. Soc. {\bf 346}, 987 \\
Zhang C. et al. 2014, Res. in Astron. and Astrophys., {\bf 14}, 1221  \\
%%%%%%%%%%%%%%%%%%%%%%%%%%%%%%%%%%%%%%%%%%%%%%%%%%%%%%%%%%%%%%%%%%%%%%%%%%%%%%%%%%%%%%%%
\end{document}